\documentclass[amsmath,amssymb,superscriptaddress,longbibliography,twocolumn]{revtex4-2}
\usepackage{graphicx}
\usepackage{varioref}
\usepackage{xr-hyper}
\usepackage{xcolor}
\usepackage{nicefrac}
\usepackage{xfrac}
\usepackage{todonotes}
\usepackage{hyperref}
\hypersetup{colorlinks,linkcolor=blue,urlcolor=teal,citecolor=orange}
\usepackage{ulem}
\usepackage{siunitx}
\usepackage{graphicx}
\usepackage{dcolumn}
\usepackage{bm}
\usepackage{xcolor}

\newcommand{\cyan}[1]{{\textcolor{black}{#1}}}

\begin{document}

\preprint{APS/123-QED}


\title{Decoupled static and dynamical charge correlations in a cuprate superconductor}

\title{Symmetry dichotomy of static and dynamical charge correlations in a cuprate superconductor}

\title{Decoupling of Static and Dynamic Charge Correlations revealed by Uniaxial Strain in a Cuprate Superconductor}

\author{L.~Martinelli}
\email{leonardo.martinelli@physik.uzh.ch}
\affiliation{Physik-Institut, Universit\"{a}t Z\"{u}rich, Winterthurerstrasse 
190, CH-8057 Z\"{u}rich, Switzerland}
\author{I.~Biało}%
\email{izabela.bialo@uzh.ch}
\affiliation{Physik-Institut, Universit\"{a}t Z\"{u}rich, Winterthurerstrasse 
190, CH-8057 Z\"{u}rich, Switzerland}%

\author{X.~Hong}%
\affiliation{Physik-Institut, Universit\"{a}t Z\"{u}rich, Winterthurerstrasse 
190, CH-8057 Z\"{u}rich, Switzerland}%

\author{J.~Oppliger}%
\affiliation{Physik-Institut, Universit\"{a}t Z\"{u}rich, Winterthurerstrasse 
190, CH-8057 Z\"{u}rich, Switzerland}%

\author{C. Lin}%
\affiliation{Physik-Institut, Universit\"{a}t Z\"{u}rich, Winterthurerstrasse 
190, CH-8057 Z\"{u}rich, Switzerland}%

\author{T. Schaller}%
\affiliation{Physik-Institut, Universit\"{a}t Z\"{u}rich, Winterthurerstrasse 
190, CH-8057 Z\"{u}rich, Switzerland}%

\author{J. Küspert}%
\affiliation{Physik-Institut, Universit\"{a}t Z\"{u}rich, Winterthurerstrasse 
190, CH-8057 Z\"{u}rich, Switzerland}%

\author{M.~H.~Fischer}
\affiliation{Physik-Institut, Universit\"{a}t Z\"{u}rich, Winterthurerstrasse 
190, CH-8057 Z\"{u}rich, Switzerland}%

\author{T.~Kurosawa}
\affiliation{Department of Physics, Hokkaido University - Sapporo 060-0810, 
Japan}
\affiliation{Department of Applied Sciences, Muroran Institute of Technology, Muroran 050-8585, Japan}

\author{N.~Momono}
\affiliation{Department of Physics, Hokkaido University - Sapporo 060-0810, 
Japan}
\affiliation{Department of Applied Sciences, Muroran Institute of Technology, Muroran 050-8585, Japan}

\author{M.~Oda}
\affiliation{Department of Physics, Hokkaido University - Sapporo 060-0810, 
Japan}

\author{D.~V.~Novikov}
\affiliation{Deutsches Elektronen-Synchrotron DESY, Notkestraße 85, 22607 Hamburg, Germany}

\author{A.~Khadiev}
\affiliation{Deutsches Elektronen-Synchrotron DESY, Notkestraße 85, 22607 Hamburg, Germany}

\author{E.~Weschke}
\affiliation{Helmholtz-Zentrum Berlin für Materialien und Energie, Albert-Einstein-Strasse 15, D-12489 Berlin, Germany.}

\author{J.~Choi}
\affiliation{Diamond Light Source, Harwell Campus, Didcot, Oxfordshire OX11 0DE, United Kingdom}
\affiliation{Department of Physics, Korea Advanced Institute of Science and Technology (KAIST), 291 Daehak-ro, Daejeon 34141, Republic of Korea}

\author{S.~Agrestini}
\affiliation{Diamond Light Source, Harwell Campus, Didcot, Oxfordshire OX11 0DE, United Kingdom}

\author{M.~Garcia-Fernandez}
\affiliation{Diamond Light Source, Harwell Campus, Didcot, Oxfordshire OX11 0DE, United Kingdom}

\author{Ke-Jin~Zhou}
\affiliation{Diamond Light Source, Harwell Campus, Didcot, Oxfordshire OX11 0DE, United Kingdom}

\author{Q.~Wang}%
\affiliation{Department of Physics, The Chinese
University of Hong Kong, Shatin, Hong Kong, China}%

\author{J.~Chang}
\email{johan.chang@physik.uzh.ch}
\affiliation{Physik-Institut, Universit\"{a}t Z\"{u}rich, Winterthurerstrasse 
190, CH-8057 Z\"{u}rich, Switzerland}%

\date{\today}

\begin{abstract}
We use uniaxial strain in combination with ultra-high-resolution Resonant Inelastic X-ray Scattering (RIXS) at the oxygen $K$- and copper $L_3$-edges to study the excitations stemming from the charge ordering wave vector in La$_{1.875}$Sr$_{0.125}$CuO$_4$.
By detwinning stripe ordering, we demonstrate that the optical phonon anomalies do not show any stripe anisotropy.
The low-energy charge 
excitations also retain an in-plane 
four-fold symmetry.
As such, we find that both phonon and charge excitations are decoupled entirely from the strength of static charge ordering. 
The almost isotropic character of charge excitations \cyan{is indicative of a quantum critical behaviour} and remains a possible source for the strange metal properties found in the normal state of cuprate superconductors. 
\end{abstract}

\maketitle

Excitation spectra 
entail fingerprints of their underlying ground state. 
In high-temperature cuprate superconductors, this has motivated numerous studies of phonon and charge excitations.
In particular, excitations with momentum matching the charge-stripe ordering wavevector have been the object of intense scrutiny \cite{arpaia2021Charge}. 
The problem is complicated by the presence of dynamic excitations \cyan{which emerge at higher temperatures and} precede the formation of static charge order \cite{ChaixNP2017, Arpaia906, MiaoPNAS2017, MiaoPRX2019, lee2021Spectroscopic, huang2021Quantum}. 
The predominant interpretation is that they constitute quantum fluctuations of the electronic ``crystal'' created \cyan{by charge order (CO)}. 
Such strong fluctuations could signal the proximity to a quantum critical point, such as
the putative pseudogap endpoint at $p\sim0.19$ \cite{arpaia2023Signaturea, caprara2022Dissipationdriven}.
It was proposed that the short-range character of such excitations could provide a quasi-isotropic scattering in momentum space, and explain the strange metal behavior in the transport properties \cite{grissonnanche2021Linearin, caprara2017dynamical, seibold2021Strange}. 
Little is known yet, however, about the symmetry of these excitations. 
The presence of twinned domains in the tetragonal CuO$_2$ planes precludes 
 investigation of their symmetry properties.

Employing a novel \textit{in-situ} strain device, we achieve substantial detwinning of the charge-stripe order in the prototypical cuprate La$_{2-x}$Sr$_x$CuO$_4$ with $x=0.125$.
We investigate the inelastic excitations using ultra-high-resolution Resonant Inelastic X-ray Scattering (RIXS) at both the oxygen-$K$ and copper-$L_3$ edge. 
No strain-induced changes in the energy and intensity of optical phonon excitations are observed
\cyan{demonstrating that the anomalies in the phonon intensity and dispersion are not primarily correlated with the strength of the stripe order parameter}. 
\cyan{The simultaneous use of ultra-high-resolution at two resonant edges resolves the apparent discrepancy observed in the RIXS response at the Cu $L_3$ and O $K$ edges \cite{ChaixNP2017,LiPNAS2020, huang2021Quantum} and provides a clear proof of the existence of quantum fluctuations associated with CO.  
Our measurements demonstrate that these excitations remain unaffected by the strain application, both around the $a-$ and $b-$ axis ordering wavevectors, and therefore retain a higher symmetry then the static order parameter. Moreover, we show that the anomalous softening of the phonon modes is a direct consequence of these collective modes. Our observations suggest that these fluctuations are linked to short-range electronic interactions, and the decoupling from the static charge-stripe order hints to a quantum critical behavior. Our measurements bear significant implications for the theory of charge fluctuations and the properties of strange metals.}

\begin{figure*}[ht]
    \centering
    \includegraphics[width=\textwidth]{Fig1.pdf}
    \caption{
    RIXS spectra at oxygen $K$-edge under uniaxial pressure application. (a) Photograph and (b) sketch of respectively the \textit{in-situ} strain device and the experimental (scattering) geometry. 
    Incident and scattered x-rays are indicated by blue wavy arrows. The reciprocal $H$ and $K$ directions are indicated with black arrows. The direction of application of uniaxial strain ($K$) is indicated with red arrows. (c)-(h) Low-energy RIXS spectra as a function of momentum along (c-e) the $(H,0)$ and (f-h) $(0,K)$ directions. (i) Intensity of static CO along $(H,0)$ (circles) and $(0,K)$ (triangles) directions as a function of applied strain. (j) \cyan{Domain population factor $\eta$ (see text for definition)} as a function of strain. \cyan{(k): Longitudinal correlation length as along the $(H,0)$ and $(0,K)$ directions as function of in-plane strain $\varepsilon_b$.}
    }
    \label{fig:fig1}
\end{figure*}
%

\textit{Methods --}
We used single-crystals of LSCO $x=0.125$ 
\cite{ChangPRB2008}. Rectangular pieces with facets defined by the tetragonal $a-$, $b-$, and $c-$axes were cut down to 1 x 0.4 x 0.7 mm$^3$ in size, glued inside the strain cell, and cleaved ex-situ.
Our uniaxial pressure cell is an adaptation of the design described in Ref.~\cite{lin2021Visualizationa,lin2024Uniaxial}. 
\cyan{Compressive strain} \cyan{is applied \textit{ex-situ} through a driving screw and gauged through a calibration curve -- see Extended Matter and Supplementary Material \cite{sm}}.
The RIXS scattering geometry is illustrated in Fig.~\ref{fig:fig1}\textcolor{blue}{(a,b)}. The sample stage can be rotated such that scans along both $q_\parallel=(H,0)$ and $(0,K)$ directions are possible under application of $b-$axis strain ($\varepsilon_b)$.
The energy resolutions at the oxygen-$K$ (copper-$L_3$ edge) were 23 (43)~meV full-width-at-half-maximum (FWHM).
Wavevectors $\textbf{q}=(q_x, q_y, q_z)$ are labelled in tetragonal reciprocal lattice units (r.l.u.) of $(2\pi/a, 2\pi/b, 2\pi/c)$, with $a=b=3.78$ \AA, and $c=13.2$ \AA. Temperature was fixed to $T_c\sim28$ K, 
to maximize the CO intensity \cite{ChangPRB2008, CroftPRB2014}. \\

\begin{figure*}[t]
    \centering
    \includegraphics[width=0.85\textwidth]{Fig2.pdf}
    \caption{
    Oxygen $K$-edge resonant x-ray scattering.  (a) Example of a fitted spectrum at Q=(0.26,0).
    (b)-(d) RIXS spectra as a function of momentum $\mathbf{q}_\parallel$ after subtraction of the elastic signal. Applied strain values and directions are indicated above the panels and in the insets, respectively. (e) Energy of the bond-stretching mode and of the low-energy mode as a function of momentum for two different values of applied strain, for both $H$ (circles) and $K$ (triangles) directions. The black dotted line is a cosine function, while the red and yellow solid lines are guides to the eye. (f)-(h) Intensity of elastic peak (empty circles) and low-energy mode (filled triangles) for both $(H,0)$ (red and orange symbols) and $(0,K)$ (blue and light-blue symbols) directions, for increasing strain application. 
    }
    \label{fig:fig2}
\end{figure*}
%
\textit{Results --} 
Oxygen $K$-edge RIXS maps along the two high-symmetry directions are shown for three pressure points in Fig.~\ref{fig:fig1}\textcolor{blue}{(c-h)}. Elastic scattering from CO is observed at $q_{co}=(\delta,0)$ and $(0,\delta)$ with $\delta\sim0.235$ r.l.u.~-- consistent with previous reports~\cite{WangNC2022,ThampyPRB2014}. 
To quantify the CO intensity versus applied strain, we model the RIXS spectra with 
an elastic peak, three inelastic excitations, and a smooth function for the electron-hole continuum (Fig.~\ref{fig:fig2}\textcolor{blue}{(a)}) -- see the Supplementary Material. 
The intensity of CO is quantified by the area of the elastic line as a function of in-plane momentum. Its analysis \cyan{as a function of strain} is shown in Fig.~\ref{fig:fig1}\textcolor{blue}{(i,j)}. Without strain application, CO is nearly equally strong along the $a-$ and $b-$axis. The minor discrepancy may originate from a small residual compressive strain induced by thermal contraction of the device.  
With application of $b-$axis strain, the $a-$axis charge order intensity $I_H$ is enhanced roughly linearly in the investigated range of strain values (Fig.~\ref{fig:fig1}(i)). Simultaneously, $b$-axis CO intensity $I_K$  decreases, 
in agreement with Ref.~\cite{ChoiPRL2022}. 
The correlation length, extracted as $\xi=\frac{1}{\text{HWHM}}$~\cite{wilkins2011comparison} (with HWHM being the half-width at half-maximum of the CO peak), slightly increases with strain along the $a$-axis, from $(9\pm1)$ to $(12\pm1)$ unit cells. No significant changes are observed along the $b$-axis. These correlation lengths are consistent with the existing literature \cite{CroftPRB2014, christensen_bulk_nodate}.
The domain population factor $\eta=2(I_H-I_K)/(I_H+I_K)$, plotted in Fig.~\ref{fig:fig1}(j), increases by more than a factor of four in the applied strain interval. 
We note that the magnitude of the reduction of CO along $b$-axis is less than the enhancement along the perpendicular direction, 
a behavior
consistent with previous measurements on YBa$_2$Cu$_3$O$_{6+x}$ \cite{KimPRL2021, kim_uniaxial_2018}. 
%
\begin{figure}[t]
    \centering
    \includegraphics[width=0.49\textwidth]{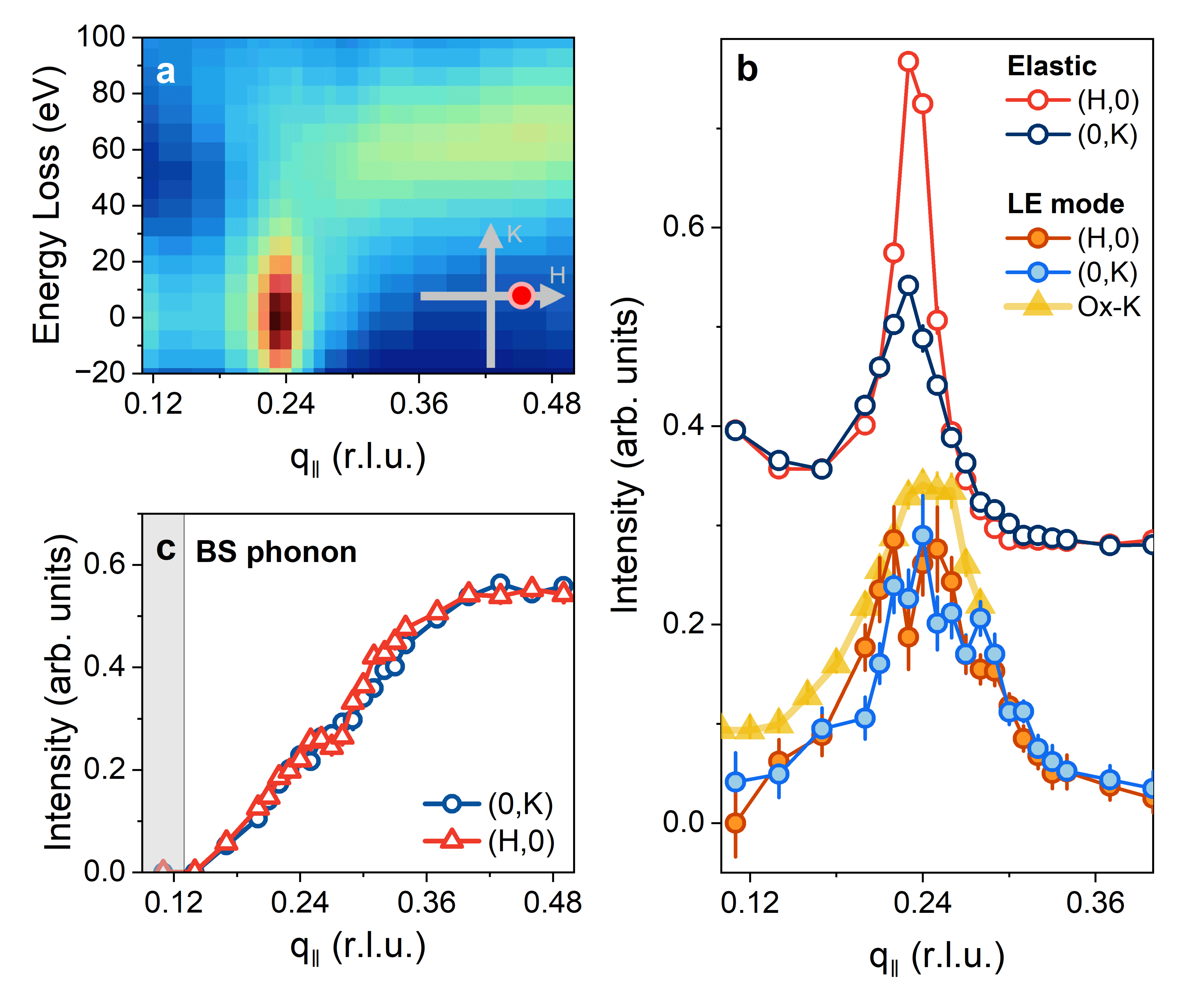}
    \caption{
    Copper $L$-edge resonant x-ray scattering.
    (a) RIXS maps at Cu $L_3$-edge along the $(H,0)$ direction at the maximum applied strain (-0.43\%). 
    (b) Comparison of the amplitude of CO fluctuations at the oxygen (yellow triangles) and copper edges (orange and light blue circles). Oxygen data has been scaled and shifted by 0.04 in the vertical direction.
    (c) Intensity of the bond-stretching phonon along the $(H,0)$ and $(0,K)$ directions at the maximum applied strain.
    }
    \label{fig:fig3}
\end{figure}

The high-energy resolution RIXS spectra
allows us to scrutinize the charge excitations visible in the $0-100$ meV energy range.
A typical RIXS spectrum near is shown in Fig.~\ref{fig:fig2}\textcolor{blue}{(a)}.
The low-energy inelastic region, visible in Fig.~\ref{fig:fig2}\textcolor{blue}{(b-d)}, contains three excitation branches -- with energy scales of about 20, 50 and 70~meV. By comparison with inelastic neutron and x-ray scattering experiments~\cite{pintschovius_anomalous_1999, mcqueeney_anomalous_1999, zhou_multiple_2005}, the two highest-energy modes are identified with the oxygen bond-buckling (BB) and bond-stretching (BS) modes. 
The nature of the low-energy (LE) excitation ($\sim20$ meV) will be discussed later.
From the fitting model (see also Supplementary Material \cite{sm}) the dispersion relations and excitation amplitudes are extracted. 
The  LE excitation has a pronounced intensity maximum centered approximately at the CO wavevector. The momentum width corresponds to a short correlation length of $\xi\sim \!6 a$. Its energy shows a minimum close to $q_{co}$ and a dispersion of $\sim\!5$ meV (Fig.~\ref{fig:fig2}\textcolor{blue}{(b))}. 
The energy of the BS phonon far from the CO wavevector has a characteristic $\cos(\pi k_{x,y})$ behavior (black dotted line in \ref{fig:fig2}(e)), and shows a gradual decrease towards the zone boundary \cite{pintschovius_anomalous_1999}. This is usually attributed to the screening provided by holes on extended orbitals \cite{falter1997Origin}. Close to $q_{co}$, it shows a marked softening of $\sim \! 10$~meV (Fig.~\ref{fig:fig2}(b,e)), in agreement with neutron scattering measurements~\cite{reznik2006Electrona,reznik2010Giant,park_evidence_2014}. 
The dispersion relations, as well as the intensities versus momentum, are -- within our instrumental sensitivity -- independent of the applied uniaxial pressure. In particular, as shown in Fig.~\ref{fig:fig2}\textcolor{blue}{(e)}, the anomalous softening of the BS phonons is independent of strain and is equal along the $(H,0)$ and $(0,K)$ directions. 
The intensity of the LE mode also shows the same behavior 
-- see Fig.~\ref{fig:fig2}\textcolor{blue}{(f)-(h)}. These observations are simultaneous with the uniaxial-strain-induced detwinning of the static CO.

\begin{figure}
    \centering
    \includegraphics[width=0.35\textwidth]{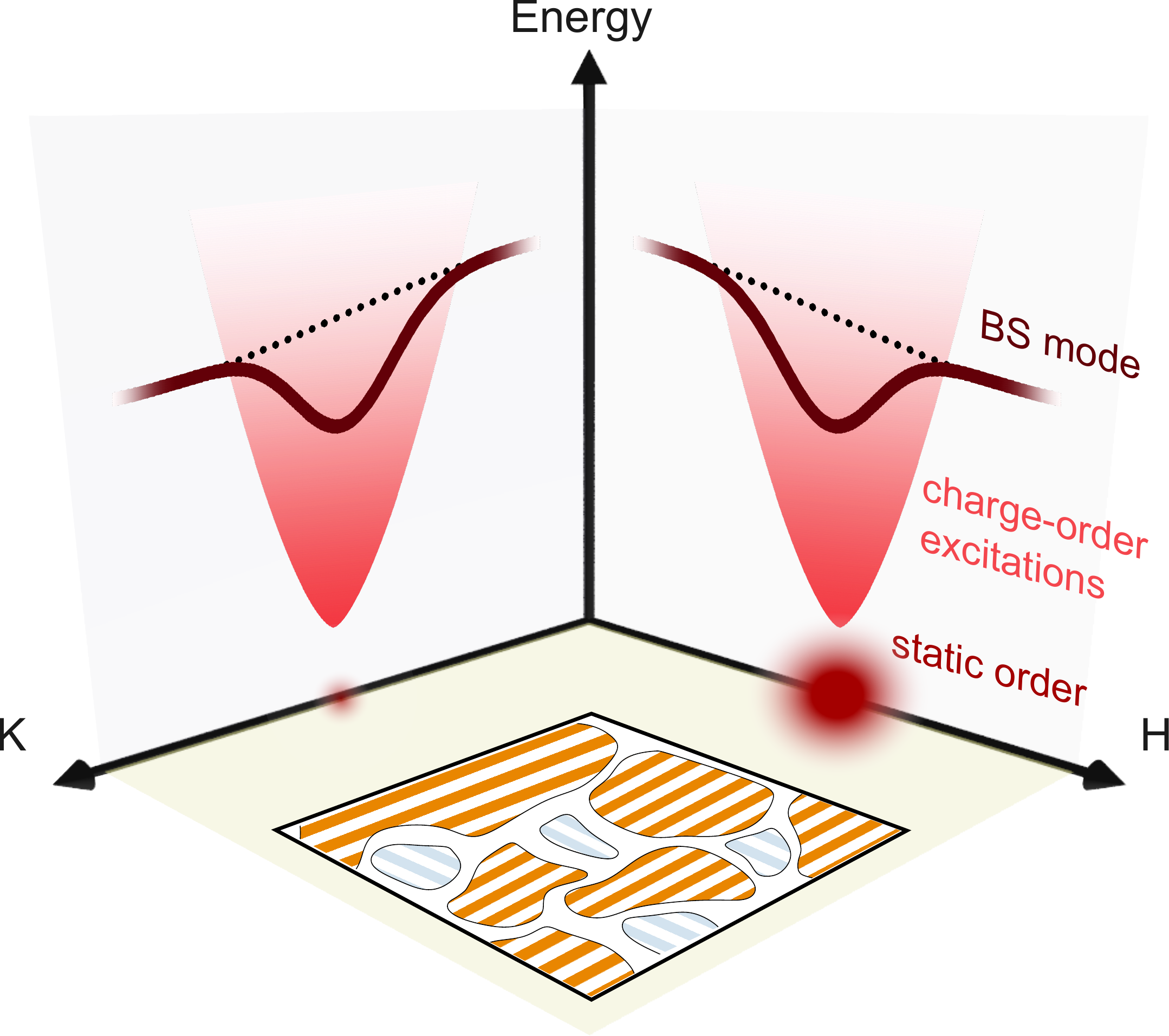}
    \caption{Sketch summarizing our experimental results. The intensity of static CO shows a substantial detwinning upon strain application. The energy and intensity of charge fluctuations, as well as the softening of the BS mode, retain tetragonal symmetry. The square at the bottom sketches the situation in real space, with an imbalance of CO domains. }
    \label{fig:fig4}
\end{figure}

These experimental observations are confirmed and reproduced at the copper $L$-edge (Fig.~\ref{fig:fig3}). 
With the assumption that the measured excitations have the same energy at the two edges, we are able to give a unified description of the RIXS response. 
\cyan{The domain population ratio $\eta$ is consistent at the two edges (Fig.~\ref{fig:fig3}a).
The $q$}-dependence of the intensity observed on the LE mode is identical to that extracted at the oxygen $K$-edge, as highlighted in panel (d) of Fig.~\ref{fig:fig3}. 
The intensity of the bond-stretching mode is, likewise, identical along the $a$- and $b$-axis along the entire  direction, and independent of strain within our experimental sensitivity (Fig.~\ref{fig:fig3}c). 
\cyan{The intensity of a phonon is, in RIXS, correlated with the electron-phonon coupling (EPC)~\cite{DevereauxPRX2016, geondzhian2018demonstration, bieniasz2021beyond}. In systems where the photo-excited electrons are strongly localized, like at the copper $L_3$-edge in cuprates, the RIXS phonon cross-section is to first order proportional to the EPC~\cite{BraicovichPRR2020}. Even in systems with mobile electrons, like at the oxygen $K$-edge, a positive correlation still persists~\cite{bieniasz2021beyond, bieniasz2022Theory}. This implies that the EPC is identical along the $a$- and $b$-axis, despite a two-fold increase in the \cyan{CO} intensity.}
These observations reinforce the results at the oxygen $K$-edge and prove that the same excitations are observed on both resonances. Their response to uniaxial strain, consistent at both absorption edges, is schematically visualized in Fig.~\ref{fig:fig4}.

\textit{Discussion.---}Acquisition of high-resolution
RIXS spectra allowed us to extract three low-energy excitations. The BB mode, at 50~meV, 
does not show anomalies in the proximity of $q_{co}$ (see Supplemental Material Fig. S4-S5), consistent with previous results.
We therefore concentrate our discussion on the other two modes.
A fundamental question revolves around the relationship between the energy of the BS mode and charge order.
The softening of the dispersion from the $\cos(\pi k_{x,y})$ behavior expected from LDA calculations \cite{giustino_small_2008} (dotted line in Fig.~\ref{fig:fig2}\textcolor{blue}{(e))}, as well as anomalies in the linewidth and intensity, have been reported in many cuprate families by several spectroscopic techniques \cite{pintschovius2004Oxygen, graf2008Bond, ChaixNP2017, LiPNAS2020, lee2021Spectroscopic}. 
\cyan{This self-energy effect could stem from an interaction between the phonon and the static order having $C_2$ symmetry, i.e. a Kohn anomaly picture \cite{KaneshitaPRL2002}.}
In this scenario, bond-stretching phonons 
should split into two branches, one propagating parallel to the stripe modulation (showing the anomalous softening) and one moving perpendicular to them \cite{KaneshitaPRL2002}. 
Without stripe-detwinning \cyan{($\eta\rightarrow0$)}, this is however difficult to verify experimentally. 
By uniaxial pressure application, we reach an almost perfect detwinning  \cyan{($\eta\rightarrow1$)}.
We do \cyan{however} not detect any change in the intensity and dispersion of phonons as a function of applied strain, despite a two-fold change in the intensity of static charge-order, see Fig.~\ref{fig:fig2}. The softening persists unaltered along both in-plane directions.
\cyan{In a Kohn-type anomaly scenario \cite{dastuto2008Sharp, maschek2015Wavevectordependent} the phonon softening is proportional to the EPC $g(q)$ and the susceptibility: $\delta \omega(q)^2 \propto g(q)^2 \chi(q,0)$ \cite{chan1973Spin}. 
\cyan{The application of $b$-axis strain increases the volume of domains ordering along the $(H,0)$ direction, and enhances (decreases) $\chi(q,0)$ ($\chi(0,q)$), as also testified by the change in correlation length (Fig. \ref{fig:fig1}k). Since our Cu $L_3$ measurements reveal that $|g|$ remains constant along 
the two Cu-O bond directions, we can conclude that a Kohn-type anomaly is not consistent with our measurements.} 
If the breathing branch were split into two branches propagating parallel and perpendicular to the stripes, the strain should modulate the intensities of the two in opposite ways along the $(H,0)$ and $(0,K)$ directions, effectively changing the centre of mass of the distribution. 
Within our sensitivity ($\sim5$ meV), we cannot detect such a shift.}

The root cause of the softening of the BS mode must therefore be searched elsewhere. Previous works suggested that dynamical fluctuations of finite energy could induce a renormalization of phonon energy \cite{reznik_giant_2010, park2014Evidence}. 
Hints of the existence of such excitations were provided only very recently in RIXS experiments on different cuprate materials.
A loss of spectral weight in the inelastic range below $\sim\!40$ meV has been reported in the low-energy spectra upon heating \cite{Arpaia906, LiPNAS2020, huang2021Quantum, arpaia2023Signaturea,Yu2020}, a behavior incompatible with a pure phonon mode. 
Similarly, when going to the strongly underdoped and overdoped regime \cite{LiPNAS2020, arpaia2023Signaturea}. 

RIXS literature on copper-$L$~\cite{ChaixNP2017, lee2021Spectroscopic} and oxygen-$K$~\cite{LiPNAS2020, huang2021Quantum} edges suggested, however, divergent interpretations. These anomalies have been assigned to either strong softening of bond-stretching modes~\cite{ChaixNP2017, lee2021Spectroscopic} or to the direct presence, in the experimental spectra, of a low-energy electronic excitation~\cite{LiPNAS2020, huang2021Quantum}. 
Both scenarios are based on the presence of a continuum of quantum fluctuations associated with charge ordering \cite{ChaixNP2017, Arpaia906, huang2021Quantum}, associated with a quantum critical point in the doping-temperature phase diagram of cuprates~\cite{caprara_dynamical_2017, seibold2021Strange}. The static order then develops out of this sea of excitations once their characteristic timescale becomes long enough.

\cyan{Within a Landau free energy paradigm \cite{Nie2014, jang2016Ideal}, the charge excitations inherit the stripe symmetry, below a characteristic energy scale $\omega_c$ linked to their effective mass. This characteristic energy $\omega_c$ in turn scales with an energetic ground-state distance to a quantum critical point. That is $\omega_c\rightarrow0$ for $x\rightarrow x_c$ where $x_c$ is the critical doping for stripe order. Excitations above the characteristic scale $\omega_c$ are expected to display critical behavior: the emergent scale invariance makes them the same on both sides of the quantum critical point \cite{kivelson2003How}.}
\cyan{We therefore assign the LE mode with the abovementioned critical excitations, in agreement with recent suggestions \cite{lee2021Spectroscopic, huang2021Quantum, arpaia2023Signaturea}.}
\cyan{The observed isotropy of the quantum fluctuations suggests that the critical energy $\omega_c$ lies well below our current experimental resolution ($\sim\!20$ meV). As such, our experiments suggest proximity to a critical point and the critical nature of the observed fluctuations. This indirect demonstration is important, as direct scaling behavior is typically smeared by doping-induced disorder. Future improvements in RIXS instrumentation may allow improved resolution that would allow access to the non-critical excitations of the stripe-ordered state.}

In conclusion, our results 
demonstrate an electronic nature of charge order beyond electron-phonon coupling. At the same time, we show that the phonon anomalies are related to a sea of dynamical fluctuations that precede the static order. The symmetry of these excitations is decoupled from static stripe order.

\bigskip 

\begin{acknowledgments}
We thank Claus Falter and S.A. Kivelson for insightful comments to the manuscript. 
L.M acknowledge support from the Swiss National Science Foundation under Spark project CRSK-2\textunderscore220797.
L.M. and I.B. acknowledge support from the Swiss Government Excellence Scholarship under project numbers ESKAS-Nr: 2023.0052 and ESKAS-Nr: 2022.0001. 
L.M. and I.B. acknowledge support from UZH Postdoc Grants project numbers FK-23-128 and FK-23-113.
J.K., L.M., and J.C. acknowledge support from the Swiss National Science Foundation under Project 200021\textunderscore188564. 
J.K. is further supported by the PhD fellowship from the German Academic Scholarship Foundation. 
Q.W. is supported by the Research Grants Council of Hong Kong (ECS No. 24306223), and the CUHK Direct Grant (No. 4053613). 
J. Choi acknowledges financial support from the National Research Foundation of Korea (NRF) funded by the Korean government (MSIT) through Sejong Science Fellowship (Grant No. RS-2023-00252768).
We acknowledge Diamond Light Source for providing beam time on beamline I21 under Proposal MM33512. We acknowledge DESY for providing beam time on beamline P23 under Proposal R-20240670 EC. We acknowledge BESSY II for providing beam time on beamline UE46-PGM1 under Proposal 232-12330-ST-1.1-P.
L.M. and I.B. have contributed equally to this work.\\
\end{acknowledgments}








\bibliography{lsco_ref}%
\clearpage
\onecolumngrid 
\begin{center}
    \textbf{End Matter} 
\end{center}

\twocolumngrid 
\cyan{\textit{Appendix A: Crystal growth} -- 
We used single-crystals of LSCO $x=0.125$ ($T_c\approx 28$~K) grown by floating zone methodology \cite{ChangPRB2008}. Information about sample characterization can be found in the Supplementary Materials \cite{sm}.
Using Laue diffraction, samples were aligned and cut with dimensions of 0.4x1x0.7~mm$^3$ along the $a$,$b$ and $c$ axes, respectively. The longest dimension coincides with the direction of strain application.
Samples were subsequently glued inside the uniaxial strain cell.} \\[1mm]

\begin{figure}
    \centering
    \includegraphics[width=0.75\linewidth]{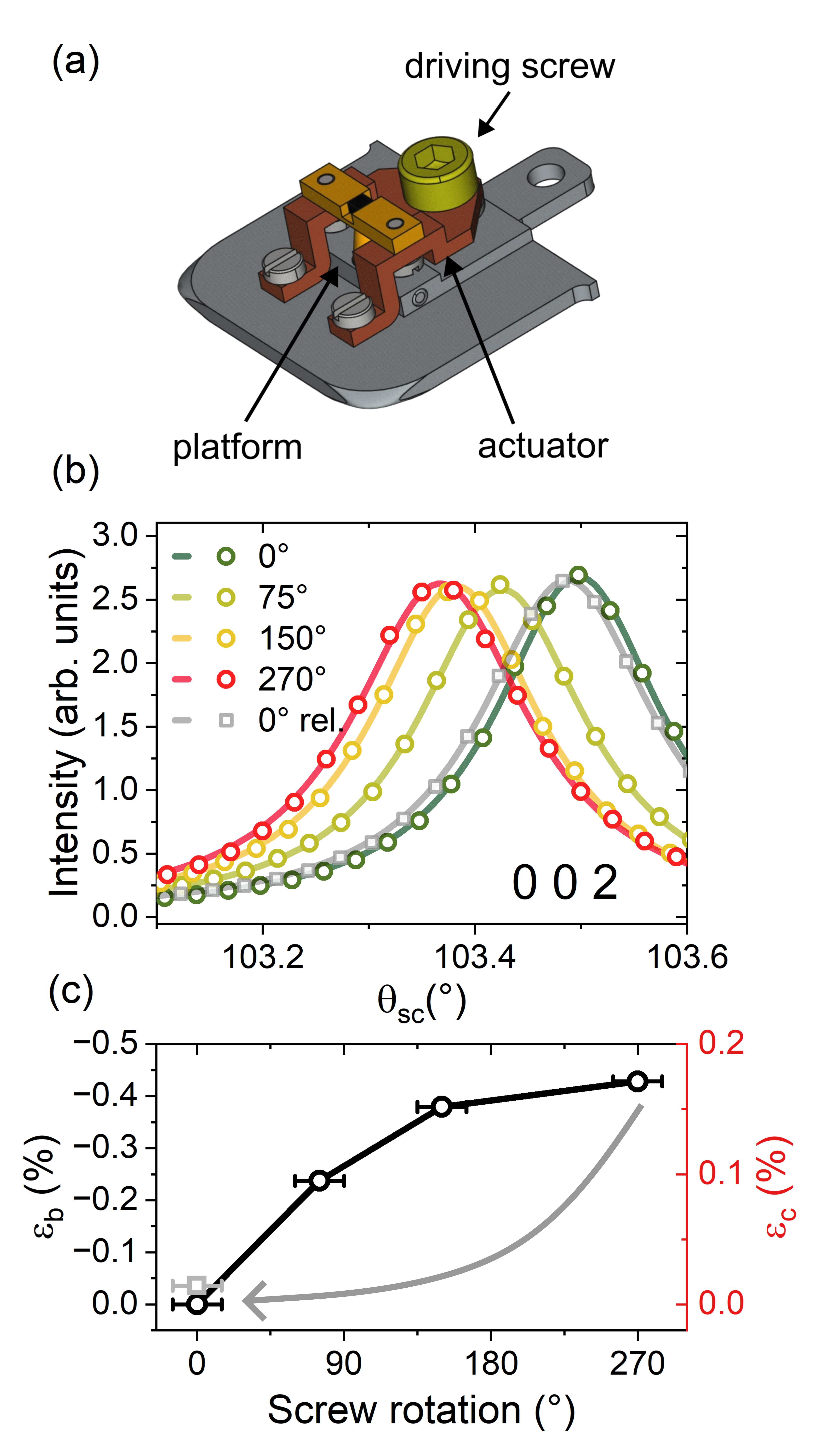}
    \caption{\cyan{Uniaxial strain device used in the RIXS measurements. (a) Sketch of the device. The platform is reported in orange, the actuator element in dark brown, and the driving screw in yellow. Sample is black. (b) $002$ reflection measured as a $\theta-2\theta$ scan as a function of driving screw rotation angle. Solid lines are Lorentzian fittings. (c) Calculated in-plane strain $\epsilon_b$ as a function of driving screw rotation angle.}}
    \label{fig:strain_device}
\end{figure}

\cyan{\textit{Appendix B: Strain cell} -- 
Our strain device see Fig.~\ref{fig:strain_device}(a)) is an adaptation of the design described in Ref.~\cite{lin2021Visualizationa}. The strain device is constructed on a base of a standard Omicron plate. The sample is mounted in the groove of a T-shaped element (the platform, shown in orange in Fig.~\ref{fig:strain_device}\textcolor{blue}{(a)}), which lays on the base of the device (gray in Fig.~\ref{fig:strain_device}\textcolor{blue}{(a)}) and on the actuator element (dark brown in the same panel). The movement of the driving screw (shown in yellow in Fig.~\ref{fig:strain_device}\textcolor{blue}{(a)}) bends the actuator up or down depending on the sense of rotation. The platform is then bent. The gap either shrinks, applying compressive strain, or opens, applying tensile strain. In particular, upon anti-clockwise screwing compressive strain is applied. The application of strain is ensured by a circlip mounted on the driving screw, below the actuator. The plate, platform, actuator and driving screw are made from BeCu, while the circlip is made of SUS-316 stainless steel. 
The calibration of the strain device (Fig.~\ref{fig:strain_device}(b-c)) has been obtained by tracking the scattering angle $\theta_{sc}$ of the $(002)$ Bragg reflection as a function of screw rotation angle. The measurements have been performed at the UE46-PGM1 beamline of BESSY II. The incident photon energy and sample temperature were fixed to respectively $1200$ eV and $30$~K. The $c$-axis of the material goes back to the bulk value upon releasing the strain (gray point and line). For more information, see Supplementary Material \cite{sm}.} \\[1mm]

\begin{figure}[hb!]
    \centering
    \includegraphics[width=0.35\textwidth]{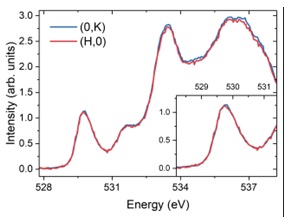}
    \caption{\cyan{Oxygen $K$-edge recorded along reciprocal $a$ and $b$-axes. The XAS spectra are essentially identical, indicating no self-absorption anisotropy.}}
    \label{fig:xas_selfabs}
\end{figure}

\cyan{\textit{Appendix C: XAS and RIXS} -- 
RIXS measurements were carried out at the I21 beamline of Diamond Light Source \cite{zhou2022I21}.
Spectra were recorded using linear vertical ($\sigma$) incident light polarization in grazing incidence scattering geometry. 
The results and conclusions presented are not influenced by self-absorption effects. First, self-absorption correction are rather small at the Zhang-Rice pre-peak of the oxygen $K$-edge. Secondly, our grazing-in configuration reduces the importance of self-absorption. Finally, and most critically, our experiments are carried out along the reciprocal $(H,0)$ and $(0,K)$ directions using identical scattering conditions (i.e.~incident and scattering angles). As such, geometry factors determining self-absorption are identical along these two high symmetry directions, so that, in principle, only differences in the self-absorption coefficient could alter the results. However, x-ray absorption spectra recorded in total-electron yield along the two directions (see figure \ref{fig:xas_selfabs}) does not show any significant difference, as expected for the small values of the applied strain.} \\[2mm]

\cyan{\textit{Appendix D: Fitting of elastic and inelastic RIXS --}
For the oxygen $K$-edge data, we fitted the spectra using a profile for the elastic line, one profile for the low-energy mode, and two phonon peaks. 
The intensity of phonon harmonics is small at the oxygen $K$edge in doped cuprates \cite{peng2022Doping}, most likely because the increased lifetime of the intermediate state (which should enhance their intensity) is compensated by a greater mobility of the photoexcited electron (which strongly reduces it) \cite{bieniasz2021beyond, bieniasz2022Theory}.
A smooth function ($k\cdot \sqrt{E_\text{loss}}$) was added to model the electron-hole continuum present at low energies. The sharp features in the spectra were modeled using the resolution lineshape. This was determined accurately by acquiring off-resonance RIXS spectra on an amorphous carbon tape.
The best fit to the experimental resolution was achieved with an asymmetric Pseudovoigt function. 
The phonons and the low-energy charge mode were fitted using the same lineshape, neglecting their intrinsic width. 
This is justified since the width of the broader phonon (the breathing mode along the $(H,0)$ direction) is at most $5-6$ meV as estimated from neutron scattering data (see e.g.~\cite{park2014Evidence} and Fig.~14 in \cite{reznik_giant_2010}).}

\cyan{Data at the Copper edge were fitted using a similar model comprising the LE mode, the buckling and breathing modes), a damped harmonic oscillator function for the paramagnon (as done routinely for doped cuprates, see e.g.~\cite{peng2018Dispersion}), and a linear function for the sum of electron-hole excitations and multimagnons. The energy of the LE mode and the two phonon modes was fixed to the one determined from the data acquired at the oxygen $K$-edge, where the higher resolution (22~meV compared to 42~meV at the copper edge) allows for a more accurate determination. 
The second harmonic of the bond-stretching mode was added \cite{BraicovichPRR2020, lee2021Spectroscopic}. The width has been fixed at two times the experimental resolution, and the energy at two times the energy of the bond-stretching mode.
More details about the fitting procedure and examples of fitted spectra at the Oxygen $K$- and copper $L$-edge are provided in the Supplementary Material \cite{sm}.} 
\\[2mm]


\end{document}